\documentclass[12pt]{article}

\usepackage{latexsym}

 \font\tenmsbm=msbm10 scaled 1200
\font\sevenmsbm=msbm9
\newfam\msbmfam
\textfont\msbmfam=\tenmsbm \scriptfont\msbmfam=\sevenmsbm

\def\msbmn{\fam\msbmfam\sevenmsbm}

\makeatletter \@addtoreset{equation}{section} \makeatother
\renewcommand{\theequation}{\thesection.\arabic{equation}}

\newcounter{parentequation}
\newenvironment{subequations}{%
  \refstepcounter{equation}%
  \begingroup
\let\protect\noexpand
  \edef\@tempa{\def\noexpand\theparentequation{\theequation}}%
  \expandafter
  \endgroup\@tempa
  \setcounter{parentequation}{\value{equation}}%
  \setcounter{equation}{0}%
  \def\theequation{\theparentequation\alph{equation}}%
  \ignorespaces
}{%
  \setcounter{equation}{\value{parentequation}}%
}
\newcommand{\eqn}[1]{(\ref{#1})}
\def\IC{\relax\,\hbox{$\inbar\kern-.3em{\rm C}$}}

\def\sIR{\hbox{\msbmn R}}
\def\sIQ{\hbox{\msbmn Q}}
\def\cA{{\cal A}} 
 \def\cD{{\cal D}}
 \def\cS{{\cal S}}
 
\def\cH{{\cal H}} 
 
\def\cL{{\cal L}} \def\cM{{\cal M}}
\def\cN{{\cal N}} 
 \def\cQ{{\cal Q}}
\def\cR{{\cal R}} \def\cV{{\cal V}}
\def\cW{{\cal W}} 
\def\beq{\begin{equation}}
\def\eeq{\end{equation}}
\def\bea{\begin{eqnarray}}
\def\eea{\end{eqnarray}}
\def\bet{\begin{tabular}}
\def\eet{\end{tabular}}
\def\bes{\begin{subequations}\bea}
\def\ees{\eea\end{subequations}}
\newcommand{\dfrac}{\displaystyle \frac}
\def\ol{\overline}

\def\a{\alpha}
\def\b{\beta}
\def\th{\vartheta}

\def\g{\gamma}
\def\G{\Gamma}
\def\s{\sigma}
\def\e{\epsilon}
\def\ve{\varepsilon}
\def\l{\lambda}
\def\z{\zeta}
\def\bz{\ol{\z}}
\def\bl{\ol{\lambda}}
\def\ta{\tilde{a}}
\def\tb{\tilde{b}}
\def\tc{\tilde{c}}
\def\td{\tilde{d}}
\def\tx{\tilde{x}}
\def\ty{\tilde{y}}
\def\tz{\tilde{z}}
\def\tw{\tilde{w}}
\def\tilt{\tilde{t}}
\def\ua{\underline{a}}
\def\ub{\underline{b}}
\def\uc{\underline{c}}

\def\tI{\tilde{I}}
\def\tJ{\tilde{J}}
\def\tK{\tilde{K}}

\begin{document}

\begin{titlepage}

\begin{flushright}
DFTT 16/2000\\
hep-th/0004111
\end{flushright}

\vspace{2cm}

\begin{center}

{\LARGE \bf General matter coupled $\cN = 2$, $D=5$ \\
gauged supergravity}

\vspace{1cm}

Anna Ceresole$^{\star}$ and Gianguido Dall'Agata$^{\sharp}$

\vspace{1cm}

{ $\star$ \it Dipartimento di Fisica, Politecnico di Torino and \\
Istituto Nazionale di Fisica Nucleare, Sezione di Torino \\
C.so Duca degli Abruzzi, 24, I-10129 Torino.\\
{\tt ceresole@athena.polito.it}}

\bigskip

{$\sharp$ \it Dipartimento di Fisica Teorica, Universit\`a  di Torino
and \\
Istituto Nazionale di Fisica Nucleare, Sezione di Torino, \\
via P. Giuria 1, I-10125 Torino.\\
{\tt dallagat@to.infn.it}}

\end{center}

\vspace{1cm}

\begin{abstract}

We give the full lagrangean and supersymmetry transformation rules
for $D=5$, $\cN=2$  supergravity interacting with an arbitrary number
of vector, tensor and hyper--multiplets, with gauging of
the R-symmetry group  $SU(2)_R$ as well as a subgroup $K$ of the
 isometries of the scalar manifold.
Among the many possible applications, this theory provides the
setting where a  supersymmetric brane--world scenario
could occur.
We comment on the presence of AdS vacua and BPS
solutions that would be relevant towards a supersymmetric smooth
realization of the Randall--Sundrum ``alternative to
compactification".
We also add some remarks on the connection between this most general
5D fully coupled supergravity model and type IIB theory on the
$T^{11}$ manifold.

\end{abstract}

\vskip 20mm


\end{titlepage}

\newpage

\baselineskip 6 mm

\section{Introduction}

In the quest for a unified description of gravity and matter interactions,
several higher dimensional theories have been proposed in the past.
In this respect, gauged supergravities, where the global isometries of the
matter lagrangean are promoted to local symmetries, have been widely explored
and by now almost all allowed models, for diverse spacetime dimensions
and  number $\cN$ of supersymmetries have been
analysed (see for instance \cite{salam,cdfbook}).

The five--dimensional $\cN = 2$ supergravity theory, in particular, has been
considered at various stages, but despite the many papers on the subject
\cite{pura}--\cite{potenz}, it still lacks a complete description
where all possible matter couplings are included and the most general
gauging is performed.

The present renewed interest in gauged supergravity
theories is mainly due to their prominent role within the
$AdS/CFT$ correspondence \cite{AdSCFT}.
It is  believed that the $5D$ gauged
supergravities provide a consistent non--linear
truncation of the lowest lying
Kaluza--Klein modes of type IIB supergravity on an $AdS_{5} \times
X^5$ space,  that should be dual to some four--dimensional
supersymmetric conformal field theory.
In detail, the truncation of the $AdS_5 \times S^5$ compactification
\cite{tronc} should be described by $\cN = 8$ gauged supergravity
\cite{N8}, while the remaining $AdS_{5} \times X^5$ models
should correspond to $\cN < 8$ gauged theories.
A special attention is devoted to the $X^5 = T^{11}$ case
\cite{Romans}, maybe the most tested non--trivial instance of
the AdS/CFT correspondence \cite{T11a,T11b}, whose
low--energy action should be expressed  in terms of an $\cN
= 2$ $U_R(1)$ gauged supergravity theory \cite{T11b}.

Still in the $AdS/CFT$ framework, gauged supergravities are very interesting
because they open the possibility of studying the renormalisation
group (RG) flows of deformations of Yang--Mills theories  by looking only
at their supergravity formulation \cite{flussusy}--\cite{KT}.
Supersymmetric \cite{flussusy} and non
supersymmetric flows \cite{flusnon,BC} to other conformal or
non--conformal theories have been studied, mostly using $\cN = 8$ gauged
supergravity.

In connection with the $\cN = 1$ CFT dual to the $AdS_5 \times
T^{11}$ compactification, the $\cN = 2$ gauged theory could be useful
to follow the RG flow associated with fractional branes
\cite{KT}, as well as a supersymmetric deformation which breaks the
$SU(2) \times SU(2)$ flavour group to the diagonal $SU(2)$, as we
will describe later.

Another line of development providing a strong motivation to find
the most general  $\cN = 2$ gauged supergravity in $D=5$
is the increasing phenomenological interest in
brane--world scenarios\footnote{
The idea of embedding our universe into an uncompactified
higher dimensional spacetime can be traced back to \cite{RT} and was further
pursued in \cite{Bandos}.}
 based on both  heterotic $M$--theory
compactifications \cite{Mth,hyper+sugra} and
Randall--Sundrum type  models \cite{RS1,RS2}.
More precisely, one can distinguish between two setups.
In the first (RS1) \cite{RS1}, meant to provide a solution to the hierarchy
problem, there are two membranes located at the orbifold fixed
points of a fifth compact dimension.
The complete action includes 5D gravity and sources for the two membranes:
$$
S = S_{bulk}+S_{branes}.
$$
In the second (RS2) \cite{RS2}, there is a single membrane source
where gravity is confined by a volcano potential given by two
surrounding AdS spaces.
Since in this case the fifth dimension is
really uncompactified, RS2 suggests an alternative to
Kaluza--Klein reduction, and one can aim at obtaining this model
within a gravity theory, without singular sources.
In order to realize RS2, one must find a model yielding
two different stable critical points with equal values of the
vacuum energy and a domain--wall solution interpolating between them.

Although it is perhaps desirable to embed any of the above
scenarios into a supersymmetric string or gravity theory,
the supersymmetrisation of the RS2 ``alternative to
compactification" is obviously much more appealing in view of its
theoretical implications.
Regrettably, Kallosh and Linde have shown \cite{KL} that none of the
available $5D$ supergravity theories allow for such RS2 construction,
and a definite answer for the minimal supersymmetric extension can only derive
from the study of the fully coupled $5D$ $\cN=2$ theory.
Notice that this is not in contrast with
the result in \cite{RSsusy}, where the RS1 scenario is considered
within the gauged $\cN = 2$ pure supergravity \cite{gaugedpura}
modified by the presence of singular sources.

Attempts to obtain any of these scenarios from a pure
stringy perspective can be found in \cite{strng}.

Aside from the  supersymmetric brane worlds, the gauged and fully coupled
$D=5$ $\cN=2$ theory is the starting point for the study of
five--dimensional black holes along the lines of \cite{BH}.

The history of the couplings and gaugings of $D=5$, $\cN = 2$
supergravity is quite long.
The pure theory was developed long ago \cite{pura}, as its
$U_R(1)$--gauged version \cite{gaugedpura}.
The interaction with vector multiplets, yielding the
general Einstein--Maxwell ungauged theory, was proposed in
\cite{MaxEin}, while some of its possible gaugings appeared in
\cite{U1g}.
As a byproduct of the above--mentioned heterotic $M$--theory
compactifications,  hypermatter was later coupled to
the abelian Einstein--Maxwell theory\footnote{The possibility of a coupling to hypermatter was
foreseen in \cite{CCDR} in the context of $M$--theory compactifications over
Calabi--Yau manifolds and a first description of such ungauged  coupling can be
found in \cite{Sierra}.} \cite{hyper+sugra}.

Very recently, the addition of tensor multiplets obtained by
dualising some of the vectors has been explored within the Einstein--Maxwell
theory, with the gauging of  a
subgroup $K\subset G$ of the isometries of the scalar manifold and
of a $U(1)$ subgroup of the
$R$--symmetry group \cite{gaugetens,potenz}.

This paper completes the above work by adding to the coupling of vector
and tensor multiplets, also interaction with an arbitrary number of
hyper--multiplets and  generic gauging of $K\subset G$ and $SU(2)_R$.
The scalar fields belonging to the vector and hyper-multiplets
parametrise a manifold $\cM$ that is the product of a very
special \cite{dWvP} by a quaternionic manifold.
Rather than the Noether method, we
use our past experience \cite{D6,D4N2} and construct the lagrangean by  a
geometrical technique that yields quite naturally also all
the higher order terms in the fermion fields, that were often neglected
in the past from both lagrangeans and supersymmetry transformation rules.
In particular, the scalar potential of the theory can be
completely expressed in terms of the shifts that appear in the
supersymmetry rules of fermionic fields due to the
gauging \cite{old}.

In sections 2--4, after a description of the generic matter couplings and $SU_R(2)$ and
Yang--Mills gaugings, we write the general lagrangean and
complete supersymmetry transformation rules.
We find that the scalar field potential is modified by the inclusion of
hypermultiplets in that a new term appears and the contribution due
to the $R$-symmetry gauging is now given in terms of $SU(2)_R$
rather than $U(1)_R$ invariant objects.

We then turn in section 5 to the investigation of the possible realization of the
RS2 scenario and perform a very simple preliminary
analysis of the scalar potential along the lines of \cite{KL}.
We seem to find that no new supersymmetric $AdS$ vacua
arise and no new BPS solutions are generated.
However, the study of possible non--BPS solutions surely deserves a deep
investigation that we postpone to future work.

As a further application of our results, we finish in section
6 with some remarks concerning the explicit realization of the theory
corresponding to the $T^{11}$ compactification of type IIB
supergravity.

\section{Preliminaries: Pure supergravity}

Rather than component formalism and Noether method, we
chose to work with superspace language for two main reasons: the
first is the use of differential forms,
that often simplify computations and make more transparent the
geometric meaning of the various structures in the theory \cite{cdfbook},
the second is the use of superfields \cite{BW}, that guarantee
a natural way to achieve supersymmetry.
The promotion of differential forms to superforms yields
the supersymmetry transformations and equations of motion, without
the need of an action to start from.
The reduction to the ordinary component formalism is trivial and
it shows that one obtains naturally all the supercovariantized
quantities.
Thus  superspace formalism is also a good tool to
simplify computations involving higher order fermion terms, which
are often disregarded in other treatments.

In order to exemplify our technique and show how to analyze the
results, we briefly revisit the pure $\cN=2$
five--dimensional supergravity \cite{pura,gaugedpura} in
the superspace formalism \cite{BW}.

The pure supergravity multiplet\footnote{Our conventions are
collected in the Appendix.}
\beq
\label{grav}
\{ e_{\mu}^a, \psi^{\a i}_\mu, A_\mu \}
\eeq contains the graviton $e_{\mu}^a$, two gravitini $\psi^{\a i}_\mu$
and a vector field $A_\mu$ (the graviphoton), that
are described in superspace by the supervielbeins
$e^{\ua} = (e^a, \psi^{\a i})$, the Lorentz connection $\omega_{\ua}{}^{\ub}$
and the one--form $A$.

The torsion $T^{\ua}$, the Lorentz and graviphoton curvatures
${R_{\ua}}^{\ub}$ and $F$ are defined by
\bea
T^{\ua} &=& D e^{\ua} = d e^{\ua} + e^{\ub} \omega_{\ub}{}^{\ua}, \nonumber\\
{R_{\ua}}^{\ub} &=& d {\omega_{\ua}}^{\ub} + {\omega_{\ua}}^{\uc}
{\omega_{\uc}}^{\ub}, \\
F &=& dA. \nonumber
\eea
These fields satisfy the  Bianchi Identities (BI)
\beq
D T^{\ua} = e^{\ub} {R_{\ub}}^{\ua}, \quad D{R_{\ua}}^{\ub} = 0, \quad dF = 0.
\eeq

It is well known that each tensor component represents a
full superfield multiplet, containing a large number of
component fields, most of which are superfluous.
The unphysical fields are eliminated by
imposing constraints on the supercurvatures.

Once the constraints are imposed, the BI of the various
superfields  are no longer automatically satisfied, and their consistent
solution determines the couplings and the dynamics of the fields,
through the derivation of the equations of motion.

Remarkably, not all the constraints
are dynamical, but some of them can be absorbed in superfield redefinitions
\cite{strategia}, that highly reduce the number of effective degrees of freedom
and simplify the solution of the BI.

Without entering the technical details, we
mention that the general strategy \cite{strategia} based on group
theoretical arguments leads to the fundamental  constraint
\beq
\label{Tfond}
T^a_{\a i \b j} = \frac{1}{2} \e_{ij} \G^a_{\a\b},
\eeq
which is needed to preserve rigid supersymmetry, and those imposing the
dynamics
\footnote{All the conventions have been chosen such that
the supersymmetry laws and the structure of the Lagrangean match
the formulae in \cite{MaxEin}.
Only the definition of the gravitational covariant
derivative differs by a sign,
that reflects in the opposite sign in the definitions of $\omega$ and $R$
with respect to \cite{MaxEin}.}:
\bea
\label{Talbega}
{T_{\a i \b j}}^{\s k} &=& 0, \\
\label{Fbase}
F_{\a i \b j} &=& - \frac{i}{4} \sqrt{6} \, \e_{ij} C_{\a\b}.
\eea
From the rheonomic approach point of view \cite{cdfbook,gaugedpura},
\eqn{Tfond} and \eqn{Fbase} are a result of the Maurer--Cartan's equations dual to
the $SU(2,2|1)$ superalgebra.

One can now solve the BI and find the following
parametrizations:
\begin{subequations}
\label{sol}
\bea
\label{torsion}
T^a &=& -\frac{1}{4} \ol{\psi}^i \g^a \psi_{i}, \\
T^{i} &=& \frac{1}{2} e^a e^b T_{ba\,i} + \frac{i}{4\sqrt{6}}
e^a \left(\g_{abc}\psi^i - 4 \eta_{ab} \g_c \psi^i\right) F^{bc}, \\
R_{ab} &=& \frac{1}{2} e^d e^c R_{cd\,ab} +\frac{1}{4} e^c
\ol{\psi}^i \g_c T_{ab\,i} - \frac{1}{2} e^c \ol{\psi}^i
\g_{[a}T_{b]c\,i}  + \nonumber \\
&+&\frac{i}{8\sqrt{6}} \left(
\ol{\psi}_i \g_{ab}{}^{cd} \psi^i + 4 \ol{\psi}_i \psi^i
\eta_{ab}^{cd}\right)F^{cd}, \\
F &=& \frac{1}{2} e^a e^b F_{ba}+ \frac{i\sqrt{6}}{8}
\ol{\psi}^i\psi_i .
\ees

This solution deserves some comments.
First of all, since the $T$-- and $F$--BI are
coupled, they cannot be solved separately, but  the ${T_{a \a
i}}^{\s k}$ component of the torsion is determined by the $F$--BI.
Although the coupling with matter
multiplets could in principle change the constraints
\eqn{Tfond}--\eqn{Fbase}  and their solution \eqn{sol},
one finds that the closure of the supersymmetry algebra still requires the
fulfillment of the fundamental constraint \eqn{Tfond}.
Moreover, it turns out that also the \eqn{torsion} solution and the
\eqn{Talbega} constraint can be preserved.
In particular, the $T^a{}_{bc}$ component of the torsion can always
be annihilated by a shift of the Lorentz connection
$\omega_{a\,b}{}^c$, translating all its non--zero components
in the definition of $T_{a\a i}{}^{\b j}$.
E.g., in our formulation, the solution given in \cite{gaugedpura}
would read
$$ T_{ab}{}^c \sim \e_{ab}{}^{cde} F_{de} $$
and
$$ T_{a\a i}{}^{\b j} \sim (\G^b)_\a{}^\b \delta_i^j F_{ab}, $$
which differs from \eqn{sol} by a shift of $\e_{abcde} F^{de}$
in the $\omega_{a\,b}{}^c$ definition.
For the \eqn{Talbega} constraint,
using the vielbein and connection redefinitions found in
\cite{strategia}, one sees that
only its $\underline{4}$ irreps.  of $SO(5)$ are physical and they
are fixed by the solution of the lowest dimensional $T$--BI which does
not change in presence of matter\footnote{We will see in the next section that
this freedom can be used to couple the hypermatter.}.
The same field redefinitions also tell that the only physical component
in $T_{a\a i}{}^b$ is the $\underline{40}$, which does not correspond
to any structure in our supergravity model and indeed
it is set to zero by the same $T$--BI.

Finally, the \eqn{sol} equations can be determined by solving only the
$F$-- and $T$--BI, since by Dragon's theorem the $R$--BI follow once
solved the $T$ ones \cite{strategia}.

The ordinary supersymmetry transformations can be easily read off
from the superspace results \eqn{sol}.
In fact, $\ve^A = (0,\ve^{\a i})$ being the translation parameter, the
supersymmetry transformations of the component fields are given by
covariantized superspace Lie derivatives of the corresponding
superfields, evaluated at $\th = 0 = d \th$:
\beq
\label{28}
\delta_{\ve}\phi = \left. \left( i_{\ve} D + D i_{\ve} \right) \phi
\right|_{\th = 0 = d\th} .
\eeq
They are explicitly
\beq
\begin{array}{rcl} \delta_{\ve}e^a &=& \dfrac{1}{2} \ol\ve^i \g^a
\psi_i ,\\
\delta_{\ve}\psi^i &=& D (\widehat{\omega})\ve_i + \dfrac{i}{4\sqrt{6}}
e^a \left(\g_{abc} - 4 \eta_{ab} \g_c \right) \ve_i \, \widehat{F}^{bc} ,\\
\delta_{\ve}{\omega_a}^b &=& \dfrac{i}{4\sqrt{6}} \left(
\ol{\psi}_i \g_{ab}{}^{cd} \ve^i + 4 \, \ol{\psi}_i \ve^i
\, \eta_{ab}^{cd}\right)\widehat{F}^{cd},\\
\delta_{\ve}A &=& \dfrac{i\sqrt{6}}{4} \ol\psi^i \ve_i\ ,
\end{array}
\eeq
where the hatted quantities refer to supercovariantized terms, i.e.
$\widehat{F}_{ab} = F_{ab} + \dfrac{i\sqrt{6}}{4} \; \ol{\psi}^i_{[a} \psi_{b]i}$.

\section{Gauged supergravity with generic matter coupling}

Five--dimensional $\cN\!=\!2$ supergravity allows,
beside the supergravity multiplet \eqn{grav},
three kinds of matter multiplets: the vector, tensor and
hypermultiplet.

The vector multiplet
$$\{A_\mu,\l^i,\phi\}$$
contains a vector field, an $SU_R(2)$
doublet of spin--$1/2$ fermions and one real scalar field, while
the tensor multiplet
$$
\{B_{\mu\nu},\l^i,\phi\}
$$
contains a tensor field of rank two,
and again an $SU_R(2)$ doublet of spin--$1/2$ fermions  and one
real scalar field.
In the hypermultiplet
$$
\{\z^A,q^X\}$$ there is a doublet of spin--$1/2$ fermions $A=1,2$ and
four real scalars $X=1,\ldots,4$.

To allow for self--interactions, the scalars of the $n_V$ vector,
$n_T$ tensors and $n_H$ hypermultiplets parametrize a manifold
$\cM_{scalar}$ which is the direct product of a very special \cite{dWvP} and a
quaternionic manifold
\beq
\cM = \cS (n_V + n_T) \otimes \cQ (n_H),
\eeq
with $dim_{\sIR} \cS = n_V + n_T$ and $dim_{\sIQ} \cQ = n_H$.

In detail, the theory  we are going to describe has the following
field content
\beq
\left\{ {e_\mu}^a, \psi^i_\mu, A_\mu^I, B_{\mu\nu}^M, \l^{i \ta}, \z^A ,
\phi^{\tx},
 q^X\right\}.
\eeq
Here $I = 0,1,\ldots ,n_V$ is an index labeling the vector fields of
the $n_V$ vector multiplets and the graviphoton, since they will mix
in the interactions.
$M = 1, \ldots, n_T$ labels the tensor multiplets.
The scalars $\phi^{\tx}$, $\tx = 1,\ldots,n_V+n_T$, parametrize the
target space $\cS$ and thus $\tx$ is a curved index.
The $\l^{\ta i}$ instead transform as vectors under the tangent
space group $SO(n_V + n_T)$ and $\ta = 1,\ldots, n_V + n_T$ is
the corresponding flat index.
The $\cQ$ manifold is the target space of the $q^X$ scalars and
 $X=1,\ldots, 4n_H$ are the curved indices labeling the
coordinates.
As expected for a quaternionic manifold, we have two types of flat
indices $A=1,\ldots,2n_H$ and $i =1,2$, corresponding to the
fundamental representations of $USp(2n_H)$ and $USp(2)\simeq SU(2)$.

We will shortly see that it is also useful to introduce a collective
index for the vector and tensor fields, which we denote by $\tI =
(I,M)$.

\subsection{The $\cS$ target space manifold}

The scalar field target manifold of the vector and tensor multiplets
$\cS$ is a very special manifold \cite{MaxEin} which can be described
by an $(n_V +n_T)$--dimensional cubic hypersurface
\beq
\label{poly}
C_{\tI\tJ\tK} h^{\tI} h^{\tJ} h^{\tK} = 1
\eeq
of an ambient space parametrized by $n_V + n_T +1$ coordinates
$h^{\tI} = h^{\tI}(\phi^{\tx})$.
It is  known that the kinetic term for such scalars can be
written in terms of the coordinates as
$$
C_{\tI\tJ\tK} h^{\tI} \partial_\mu h^{\tJ} \partial^\mu h^{\tK},
$$
where $C_{\tI\tJ\tK}$ is a completely symmetric {\it constant} tensor
that determines also the Chern--Simons couplings of the vector fields.

A complete classification of the allowed homogeneous manifolds
has been given in \cite{dWvP} and a lot of their
interesting properties, especially when they are restricted to be a
coset of the Jordan family, have been given in \cite{MaxEin}, to which we
surely refer the reader for all the details.

Here we only collect some notions and results for their geometrical
structures, which are directly related to the computations we present
in this and the next sections.

First, we note that when $n_T \neq 0$ not all the $C_{\tI \tJ \tK}$
coefficients differ from zero, but, as stated in \cite{gaugetens}, the
only components that survive the gauging of a Yang--Mills group are
the $C_{IJK}$ and $C_{IMN}$.

For what concerns the $\cS$ manifold, $f_{\tx}^{\ta}$, $g_{\tx\ty}$
and $\Omega_{\tx}^{\ta\tb}$ denote its $(n_V+n_T)$--bein, the metric
and the spin connection, which can be given implicitly in terms of
$f_{\tx}^{\ta}$ through the formula
$$
f_{[\tx,\ty]}^{\ta} + \Omega_{[\ty}^{\ta\tb} f_{\tx]}^{\tb} = 0.
$$
The Riemann tensor is given by
\beq
\label{sR}
K_{\tx\ty\tz\tw} = \frac{4}{3} \left( g_{\tx[\tw} g_{\tz]\ty} +
T_{\tx[\tw}{}^{\tilt} T_{\tz]\ty\tilt}\right),
\eeq
where $T_{\tx\ty\tz}$ is a completely symmetric function of
$\phi^{\tx}$.
The coordinates of the ambient space $h^{\tI}$ have an index which is
raised and lowered through the $\phi^{\tx}$--dependent metric
$a_{\tI\tJ}$.

All these functions are subject to the following algebraic and differential
constraints, that are essential to close the supersymmetry
algebra \cite{MaxEin}:
\bea
C_{\tI\tJ\tK} &=& \frac{5}{2} h_{\tI} h_{\tJ} h_{\tK} - \frac{3}{2}
a_{(\tI\tJ} h_{\tK)} + T_{\tx\ty\tz} h^{\tx}_{\tI} h^{\ty}_{\tJ}
h^{\tz}_{\tK}, \nonumber \\
h^{\tI} h_{\tI}^{\tx} &=&0, \qquad h_{\tx}^{\tI} h_{\ty}^{\tJ}
a_{\tI\tJ} =
g_{\tx\ty}, \nonumber \\
\label{35}
a_{\tI\tJ} &=& h_{\tI} h_{\tJ} + h^{\tx}_{\tI} h^{\ty}_{\tJ}
g_{\tx\ty}, \\
h_{\tI,\tx} &=& \sqrt{\frac{2}{3}} h_{\tI\tx}, \quad h^{\tI}{}_{,\tx}
=
- \sqrt{\frac{2}{3}} h^{\tI}_{\tx}, \nonumber \\
h_{\tI\tx;\ty} &=& \sqrt{\frac{2}{3}} \left(g_{\tx\ty} h_{\tI} +
T_{\tx\ty\tz} h^{\tz}_{\tI}\right), \quad h^{\tI}{}_{\tx;\ty} = -
\sqrt{\frac{2}{3}} \left(g_{\tx\ty} h^{\tI} +
T_{\tx\ty\tz} h^{\tz\,\tI}\right). \nonumber
\eea

\subsection{The $\cQ$ target manifold}

Self--interacting hypermultiplets in an $\cN =2$, $D =
5$
theory are known to live on a quaternionic K\"ahler manifold
\cite{BaggWitt,Sierra,hyper+sugra}.
The quaternionic metric tensor will be denoted by $g_{XY}(q)$, while
${\omega_{X\,i}}^j (q)$ and ${\omega_{X\,A}}^B (q)$ will be the
$USp(2)$ and $USp(2n_H)$ connections.

As a consequence of its quaternionic structure, the holonomy group of
the  manifold $\cQ$ is a direct product of $SU(2)$ and some
subgroup of the symplectic group in $2n_H$ dimensions.
This means that one can introduce the vielbeins $f_{iA}^X$ to pass to
the flat indices $iA \in USp(2) \otimes USp(2n_H)$.

These vielbeins obey the following relations:
\bea
&& g_{XY} \, f^X_{iA} \, f^Y_{jB} = \e_{ij} \, C_{AB}, \nonumber \\
&& f^X_{iC} \, f^{YC}_j + f^Y_{iC} \, f^{XC}_j = g^{XY} \, \e_{ij}, \\
&& f^X_{iA} \, f^{Yi}_B + f^Y_{iA} \, f^{Xi}_B = \frac{1}{n_H} \,
g^{XY}\,
C_{AB}, \nonumber
\eea
where $\e_{ij}$ and $C_{AB}$ are the $SU(2)$ and $USp(2n_H)$
invariant
tensors respectively.

To obtain the coupling of these fields to supergravity when they are
chargeless with respect to the Yang--Mills (Maxwell) fields, one uses
the freedom left by the solution of the $T$--BI of introducing some
fermions in ${T_{\a i\b j}}^{\g k}$, as explained in the analogous
six--dimensional case \cite{D6}.

Since the structure of the possible spinor component ${T_{\a i\b
j}}^{\g k}$ is of the same form as the pullback of the $USp(2)$
connection on the cotangent bundle basis of the superspace, one can
write the new constraint
\beq
{T_{\a i\b j }}^{\g k} = \delta_{\a}^{\g} \omega_{\b j\,i}{}^k + (\a i
\leftrightarrow \b j),
\eeq
and solve the BI, but this of course breaks $USp(2)$ covariance.
It is more convenient to proceed in a slightly different but
equivalent way which does not break covariance.

Redefining the covariant derivative $D$ by introducing the
$USp(2) \otimes USp(2n_H)$ connections
\beq
D = d + \omega_{Lorentz} + d q^X\, \omega_{X\#}{}^\#, \qquad
\# = \{A,i\},
\eeq
one gets the new torsion definitions
\beq
\begin{array}{rcl}
T^a &\equiv& d e^a + e^b {\omega_b}^a, \\
T^{\a i }&\equiv& d \psi^{\a i} + \psi^{\b i} {\omega_\b}^\a +
\psi^{\a j} {\omega_j}^i,
\end{array}
\eeq
and imposes again the constraint $ {T_{\a i\b j}}^{\g k} = 0$.
This modifies also the $T$--BI according to
\beq
D_{(new)} T^{\a i} = \psi^{\b i} {R_\b}^\a + \psi^{\a j} {\cR_j}^i,
\eeq
where $ {\cR_j}^i \equiv d {\omega_j}^i + {\omega_j}^k{\omega_k}^i
= \frac{1}{2} dq^X dq^Y \cR_{YX\,j}{}^i$ is the $USp(2)$ curvature.

It is known \cite{BaggWitt} that a quaternionic manifold is
maximally symmetric.
This fixes the $USp(2)$ curvature to
\beq
\cR_{XY\,ij} = \kappa \left( f_{XiC} f^{C}_{Yj} - f_{YiC}
f^{C}_{Xj}\right),
\eeq
with $\kappa$ a constant fixed by supersymmetry
requirements to $\kappa = -1$.

Other useful identities regarding the definition of the $\cQ$--Riemann
tensor and the $USp(2n_H)$ curvature are
\bea
\label{usp2n}
\cR_{XY\,AB} &=& \kappa \left( f_{XiA} f^i_{YB} - f_{YiA}
f^i_{XB}\right) + f_X^{iC} f_{Yi}^D \Omega_{ABCD},\\
\cR_{XY \, WZ} \; f^W_{iA} f^Z_{jB} &=& \e_{ij} \cR_{XY\,AB} + C_{AB}
\cR_{XY\,ij},
\eea
that can easily be pulled back on superspace.
Here $\Omega_{ABCD}$ denotes the totally symmetric tensor of $USp(2n_H)$.

\subsection{The Gauging}

The gauging of matter coupled $\cN = 2$
supergravity theories is achieved by identifying  the gauge group $K$ as a
subgroup of the isometries $G$ of the $\cM$  product space.
If one choses to gauge $n_V+1$ vector fields, one is left with up to
$n_T = \hbox{dim} G-n_V$ other ones, charged under $K$,
which will be dualised to tensor fields.
As explained in \cite{D4N2}, two main cases can occur: $K$ non
abelian and $K = U(1)^{n_V+1}$.
In the first case, supersymmetry requires $K$  to be a subgroup of the
full $\cM$, and the hypermultiplet space will generically split into
$$
n_H = \sum_i n_i R_i + \frac{1}{2} \sum_l n_l^P R_l^P, $$ where $R_i$
and $R_l^P$ are a set of irreps of $K$ ($P =$ pseudoreal).
In the abelian case, the $\cS$--manifold is not required to have any
isometry and if the hypermultiplets are charged with respect to the
$n_V + 1$ $U(1)$'s, the $\cQ$ manifold should at least have $n_V + 1$
abelian isometries.

The gauging now proceeds by introducing $n_V +1$ Killing vectors
acting generically on $\cM$:
\bea
\phi^{\tx} &\to& \phi^{\tx} + \e^I K_I^{\tx}(\phi), \nonumber\\
q^X &\to& q^X + \e^I K_I^X (q),\nonumber
\eea
for an infinitesimal parameter $\e^I$.

The quaternionic structure of $\cQ$ implies that $K_I^X$ can be
determined in terms of the Killing prepotential $P_{I\,i}{}^j(q)$
\cite{D4N2}, which satisfies
\beq
\label{prep1}
\cD q^Y K^X_I \cR_{XY i}{}^j = \cD P_{I\,i}{}^j
\eeq
and, for $f_{IJ}^K$ the gauge group structure constants,
\beq
\label{prep2}
g \, \cR_{XY\,i}{}^j K^X_I K^Y_J + g_R P_{[I\,i}{}^k P_{J]\,k}{}^j + g \,
\frac{1}{2}\,  f^K_{IJ} P_{K\,i}{}^j = 0.
\eeq

Gauging the supergravity theory is now done by gauging the composite
connections of the underlying $\s$--model.
Following the well--established general procedure \cite{salam,cdfbook}
first introduced in \cite{dWN} for the
$\cN =8$, $D = 4$ case, we proceed by replacing
for the YM couplings the covariant derivatives on the
scalar and fermion fields containing the Lorentz, $SO(n_V + n_T)$ and $USp(2)
\otimes USp(2n_H)$ connections by $K$--covariant derivatives
\beq
\begin{array}{rcl}
\cD \phi^{\tx} &=& D \phi^{\tx} + g A^I K_I^{\tx} (\phi),
\phantom{\dfrac{A}{A}}\\
\cD q^X &=& D q^X + g A^I K_I^X (q), \phantom{\dfrac{A}{A}}\\
\cD \l^{\ta}_i &=& D \l^{\ta}_i+ g  A^I L_I{}^{\ta\tb} (\phi)
\l^{\tb}_i, \phantom{\dfrac{A}{A}}\\
\cD \z^A &=& D \z^A + g  A^I \omega_{I\,B}{}^A (q)
\z^B,\phantom{\dfrac{A}{A}}
\end{array}
\eeq
where $g $ is the coupling constant, $A^I$ the gauge field
one--forms, $L_I^{\ta\tb}$ the $G$--transformation matrices of the
gluinos 
\beq
L_I^{\ta\tb} \equiv \partial^{\tb} K_I^{\ta}
\eeq
and
\beq
\omega_{I\,B}{}^A \equiv K_{IX;Y} f^{XA}_i f^{Yi}_B.
\eeq
At the same time, the connection $\Omega_{\ta\tb}$ is replaced by its
gauged counterpart
\beq
\cD \phi^{\tx} \Omega_{\tx\,\ta\tb} + g  A^I K_{I\ta;\tb}.
\eeq
This reflects into suitable changes in the definition of the gauged curvatures and
BI:
\bea
\cD^2 \phi^{\tx} &=& g  F^I K_I^{\tx}, \\
\cD^2 q^X &=& g  F^I K_I^X ,\\
\label{D2l}
\cD^2 \l^{\ta i} &=& {\cR_j}^i \l^{\ta j} + K^{\ta\tb} \l^{\tb i} +
g  F^I L_I{}^{\ta\tb} \l^{\tb i}, \\
\cD^2 \z^A &=& \cR_B{}^A \z^B + g  F^I \omega_{I\,B}{}^A \z^B,
\eea
where
\bea
K_{\ta\tb} &=& \frac{1}{2} \cD \phi^{\tx} \cD \phi^{\ty} K_{\ty\tx\,
\ta\tb}, \\
\cR_{AB} &=& \frac{1}{2} \cD q^X \cD q^Y \cR_{YX\,AB},
\eea
are the $\cS$ Riemann and $USp(2n_H)$ curvatures, with components
defined in \eqn{sR}, \eqn{usp2n} and
\beq
F^I = d A^I - \frac{1}{2} g  f^I_{JK} A^J A^K
\eeq
is the gauge field strength satisfying the BI
\beq
\cD F^I = 0.
\eeq

For the $SU_R(2)$ connection, the existence of a Killing vector
prepotential ${P_{I\,i}}^j(q)$ allows the following definition:
\beq
\omega_i{}^j \to \omega_i{}^j  + g_R A^I P_{I\,i}{}^j(q),
\eeq
which replaces the $SU(2)_R$ connection with its gauged
counterpart.
This implies that the new covariant derivative acting on the gravitino is
\beq
T^{\a i} \equiv \cD \psi^{\a i} =  d \psi^{\a i} + \psi^{\b i} {\omega_\b}^\a +
\psi^{\a j} {\omega_j}^i + g_R \, \psi^{\a j} \, A^I
P_{I\,j}{}^i(q),
\eeq
and that the $SU_R(2)$ curvature definition is replaced by
\beq
\widehat{\cR}_i{}^j = \frac{1}{2} \cD q^X \cD q^Y \cR_{YX\,i}{}^j + g_R
F^I P_{I\,i}{}^j ,
\eeq
provided $P_{I\,i}{}^j $ satisfies the \eqn{prep1} and \eqn{prep2}
relations.

This leads to a further redefinition of the gaugino BI \eqn{D2l} and
of the torsion one, which now read
\beq
\cD^2 \l^{\ta i} = \widehat{\cR}_j{}^i \l^{\ta j} + K^{\ta\tb}
\l^{\tb i} + g  F^I L_I{}^{\ta\tb} \l^{\tb i}
\eeq
and
\beq
\cD T^{\a i} = \psi^{\b i} R_\b{}^\a + \psi^{\a j}
\widehat{\cR}_j{}^i.
\eeq

\subsection{The solution of the Bianchi Identities}

In order to solve the superspace BI in presence of gauging
and to obtain the new susy rules, one must
first face a technical problem, that is how to
implement the tensor multiplet BI in superspace.

It is known \cite{gaugetens} that the $B^{M}_{\mu\nu}$ fields
satisfy a first order equation of motion of the type
\beq
\label{firsto}
\cD B^M = \cM^M{}_N \star B^N,
\eeq
where $\cM$ is a mass matrix.
The problem is that \eqn{firsto} contains the Hodge star product,
which is not well defined in superspace.
Moreover, being a first order equation of motion, \eqn{firsto}
cannot be derived by the standard procedure of solving the BI for the
superfield two--form $B^M$, after imposing some constraint
on its field strength $H^M$.

It must also be noted that the $B^M$ transform under the gauge group
$K$ and the correct definition of $H^M$ is
\cite{gaugetens}
\beq
H^M \equiv dB^M + g  \Lambda^M_{IN} A^I B^N,
\eeq
where $\Lambda^M_{IN}$ is the representation matrix\footnote{It has
been shown \cite{gaugetens} that they must lie in a symplectic
representation of the gauge group $K$.}.

This implies that the $H$--BI become
\beq
\cD H^M = g  \Lambda^M_{IN} F^I B^N
\eeq
where the superfield connection $B^M$ appears explicitly,
breaking covariance.

The solution to this problem lies in the origin of the tensor fields.

In perfect analogy with the six--dimensional case \cite{d6g}, before
the gauging the $B^M$ degrees of freedom are described by
$A^M$ vectors, transforming nontrivially under $K$.
To obtain the gauging of $K$, one must also introduce some
$b^M$ tensor fields with the usual invariance $\delta b^M = d\Lambda^M$.
These must be introduced in the lagrangean and transformation rules
in such a way that everywhere $F^M$ gets replaced by the combination
\beq
\label{bF}
B^M = b^M + F^M.
\eeq
Closure of the supersymmetry algebra imposes that the vector
field $A^M$ transforms non--trivially under the gauge transformations of
the tensor fields
\beq
\delta b^M = d \Lambda^M \Rightarrow \delta A^M = -\Lambda^M,
\eeq
leaving the \eqn{bF} combination gauge invariant.

This allows the vector fields to be completely gauged away by taking
$\Lambda^M = A^M$, leaving $B^M = b^M$, where now $b^M$ has ``eaten''
the  $A^M$ degrees of freedom, and
obtained its longitudinal modes by a Higgs--type mechanism.
This also implies that $B^M$ is now massive, and its supersymmetry
variation acquires an additional term of the form $d \delta_{susy}
A^M$.

From the superspace point of view, this can be seen as the need of
imposing directly on $B^M$ the same constraints and $F$--BI solutions
imposed on $F^M$ (which is now allowed, since the massive $B^M$
has lost the gauge invariance under $\delta B^M = d \Lambda$) and then
solve consistently the $H$--BI.
These BI now will also provide the first--order $B^M$ equations of
motion at the level of the $ab \a i \b j$ sector.

Following \cite{gaugetens}, from now on we will use $\cH^{\tI}$ to
denote collectively the $F^I$ and $B^M$ fields, depending on the
value of $\tI$.

\bigskip

The constraints to be imposed on the curvatures
are the straightforward generalization of those proposed for
pure supergravity \eqn{Tfond}--\eqn{Fbase}, and read
\begin{subequations}
\label{vincoli}
\bea
T^a_{\a i \b j} &=& \frac{1}{2} \e_{ij} \G^a_{\a\b}, \\
\label{Fab}
\cH^{\tI}_{\a i \b j} &=& -\frac{\sqrt{6}}{4} i \e_{ij} C_{\a\b}
h^{\tI}, \\
T_{\a i \b j}{}^{\g k} &=& 0.
\ees
In addition, one must fix the normalization of the fermion fields:
\bea
\label{Df}
\cD_{\a i} \phi^{\tx} &\equiv& - \frac{i}{2} f^{\tx}_{\ta}
\l^{\ta}_{\a i }, \\
\label{Dq}
\cD_{\a i} q^X &\equiv& i  f_{Ai}^X \z_\a^A.
\eea
Introducing the \eqn{vincoli} constraints and the \eqn{Df}--\eqn{Dq}
definitions in the BI, we obtain the new parametrizations of the
curvatures as well as some algebraic and differential constraints on
the geometric structures (e.g.  those presented in \eqn{35}) and the
equations of motion.

The torsion parametrization remains the same  as in the
pure gravity case:
\beq
\label{fondvin}
T^a = - \frac{1}{4} \ol{\psi}^i \g^a \psi_i.
\eeq
However,
the super--field--strength of the gravitino now contains many new
terms
involving couplings with fermions and a new gauging term,
which
is fixed by the solution  of the $T$--BI:
\bea
T_i &=& \frac{1}{2} e^a e^b T_{ba\,i} + \frac{i}{4\sqrt{6}} h_{\tI}
e^a \left(
\g_{abc}\psi_i - 4 \eta_{ab} \g_c \psi_i\right) \cH^{bc\,\tI} +
\nonumber \\
&-& \frac{1}{12} e^a \g_{ab} \psi^j \; \bl^{\tc}_i \g^b
\l_j^{\tc} + \frac{1}{48} e^a \g_{abc} \psi^j \;
\bl_i^{\ta}
\g^{bc} \l_j^{\ta} + \frac{1}{6} e^a \psi^j \;
\bl_i^{\tc} \g_a
\l_j^{\tc}  + \nonumber \\
&-& \frac{1}{12} e^a \g^b\psi^j \; \bl_i^{\tc} \g_{ab}
\l_j^{\tc}
 + \frac{1}{8} e^a \g^{bc} \psi_i \; \bz_A \g_{abc}
\z^A +\frac{i}{\sqrt{6}} g_R \, e^a \g_a \psi^j P_{ij},
\eea
where
\beq
\label{Pij}
P_{ij} \equiv h^I P_{I ij}.
\eeq

Regarding the Yang--Mills and tensor multiplets, the scalar
parametrization directly follows from \eqn{Df}:
\beq
\cD \phi^{\tx} = e^a \cD_a \phi^{\tx} + \frac{i}{2} \ol{\psi}^i
\l_i^{\ta} \, f_{\ta}^{\tx}.
\eeq

The $F^I$ field--strength, apart from the component fixed in
\eqn{Fab}, has a new term involving the gluino fields which is fixed
by the $F$--BI.
Since the $B^M$ parametrization must have the same form, these can be
collectively written as
\beq
\cH^{\tI} = \frac{1}{2} e^a e^b \cH_{ba}^{\tI} - \frac{1}{2} e^a
\; \ol{\psi}^i \g_a \l_i^{\ta} \, h_{\ta}^{\tI} +
\frac{i\sqrt{6}}{8} \ol{\psi}^i\psi_i \, h^{\tI}.
\eeq

The gluinos field strength (and  supersymmetry variation)
are
\bea
\cD\l_i^{\ta} &=& e^a \cD_a \l_i^{\ta} - \frac{i}{2} f^{\ta}_{\tx}
\g^a \psi_i
\, \cD_a \phi^{\tx} +\frac{1}{4} h_{\tI}^{\ta} \g^{ab} \psi_i \,
\cH_{ab}^{\tI}  + \nonumber \\
\label{Dlam}
&-& \frac{i}{4\sqrt{6}} {T^{\ta}}_{\tb\tc} \left[ -3 \psi^j \,
\left(\bl^{\tb}_i \l_j^{\tc}\right) + \g_a\psi^j \,
\left(\bl^{\tb}_i \g^a \l_j^{\tc}\right) + \frac{1}{2} \g_{ab} \psi^j
\, \left(\bl^{\tb}_i \g^{ab} \l_j^{\tc}\right)\right]+ \\
&+& g_R \psi^j P^{\ta}_{ij} + g  W^{\ta} \psi_i,  \nonumber
\eea
where
\beq
\label{Paij}
P^{\ta}_{ij} \equiv h^{\ta I} P_{I\, ij}
\eeq
and
\beq
\label{Wa}
W^{\ta} \equiv -\frac{\sqrt{6}}{8} \Omega^{MN} h_M^{\ta} h_N .
\eeq
The \eqn{Dlam} parametrization contains the obvious terms needed to
close the $F$--BI and the supersymmetry algebra on $\phi^{\tx}$.
In addition, the first bilinear in the gluini and the $R$--symmetry
gauging terms are fixed by closure of the supersymmetry algebra.
The Yang--Mills gauging term is then fixed by the $H$--BI, since it
appears only when tensor multiplets are involved \cite{gaugetens}.

We point out that, differently from the four--dimensional case,
the Yang--Mills gauging in the absence
of tensor multiplets does not lead to any extra term.
This is due to the fact that, according to \cite{U1g}, $h^I K_I^{\tx}
= 0$ if we assume that under an infinitesimal $K$--transformation
with parameter $\eta$, the vectors transform as
\beq
\delta_\eta A^I \sim f^I_{JK} A^J \eta^K.
\eeq
When the tensor fields are involved, the Killing vectors get a new
contribution coming from the $K$--transformation properties of the tensors
and therefore now $h^I K_I^{\tx} \neq 0$.
This gives us back a Yang--Mills gauging term, determined by closure of
the supersymmetry algebra on $\phi^{\tx}$, which must be
equivalent to \eqn{Wa} and it reads
\beq
W^{\ta} = \frac{\sqrt{6}}{4} h^I K_I^{\tx} f_{\tx}^{\ta}.
\eeq

\medskip

The solution of the $H$--BI fixes  the $H^M$
parametrization, which reads
\beq
H^M = \frac{1}{3!} e^a e^b e^c H_{cba} + \frac{i}{8} g  e^b e^a
\, \ol{\psi}^i \g_{ab} \l_i^{\ta} h_{N} \Omega^{MN} +
g
\frac{\sqrt{6}}{16} e^a \, \ol{\psi}^i \g_a \psi_i \,
\Omega^{MN} h_N,
\eeq
where it must be noted that in $H_{abc}$ one should substitute
the $B$ equations of motion.

The closure of the $H$--BI also imposes some constraints on
the representation matrix $\Lambda$:
\beq
\begin{array}{rcl}
\Omega^{MN} h_N &=& \sqrt{6} \Lambda^M_{IN} h^I h^N ,
\phantom{\dfrac{A}{A}}\\
\Omega^{MN} h^{\ta}_N &=& \sqrt{6} \Lambda^M_{IN} \left(h^{\ta I}
h^N +  h^I h^{\ta N} \right) ,
\end{array}
\eeq
which can be shown to be equivalent to condition (5.12) of
\cite{gaugetens}, namely
\beq
\Lambda^M_{IN} = \frac{2}{\sqrt{6}} \Omega^{MP} C_{NPI}.
\eeq

Turning to the hypermultiplets, the scalar parametrization is
fixed once \eqn{Dq} is imposed, and is given by
\beq
\cD q^X = e^a \cD_a q^X - i \ol{\psi}^i \z^A \, f_{iA}^X.
\eeq
Closure of supersymmetry on $q^X$ and $\z^A$ imposes then
\beq
\cD \z^A = e^a \cD_a \z^A - \frac{i}{2}\g_a \psi^i \cD_a q^X \,
f_{iX}^A +  g  \psi^i \cN_{i}^A,
\eeq
with the $g $--order shift defined as
\beq
\label{NiA}
\cN_{iA} \equiv \frac{\sqrt{6}}{4} f_{XAi} K^X_{I} h^{I}.
\eeq
This term is due to the Yang--Mills charge of the
hypermultiplets.

Finally, for completeness, we give here also the parametrization of
the Riemann tensor:
\bea
R_{ab} &=& \frac{1}{2} e^d e^c R_{cd,ab} + \frac{i}{8\sqrt{6}} \left(
\ol{\psi}_i \g_{ab}{}^{cd} \psi^i + 4 \ol{\psi}_i \psi^i
\eta_{ab}^{cd}\right)\cH^{\tI}_{cd} h_{\tI} + \nonumber \\
&+& \frac{1}{24} \ol{\psi}^i \g_{abc} \psi^j \, \bl_i \g^c \l_j +
\frac{1}{48} \ol{\psi}^i \g_{c} \psi^j \, \bl_i \g^{abc} \l_j +
\nonumber
\\
&+& \frac{1}{24} \ol{\psi}^i \psi^j \, \bl_i \g_{ab} \l_j +
\frac{1}{4}
\ol{\psi}_i \g^c \psi_j \; \z_A \g_{abc} \z^A
-\frac{i}{2\sqrt{6}}
\ol{\psi}^i \g_{ab} \psi^j \, P_{ij} + \nonumber \\
&+& \frac{1}{4} e^c \ol{\psi}^i \g_c T_{ab\,i} - \frac{1}{2} e^c
\ol{\psi}^i
\g_{[a}
T_{b]c\,i} .
\eea

As shown in the previous section,
the ordinary supersymmetry transformations of the fields
are recovered from the above superspace results by using \eqn{28}.
One sees that, as expected,
they complete those given in \cite{gaugetens} with the higher
order Fermi terms and the new couplings:
\bea
\delta_{\ve} e^a &=&
\frac{1}{2} \bar{\ve}^i \g^a \psi_i,\\
\delta_{\ve} \psi_i &=&
\cD(\widehat{\omega}) \ve_i + \frac{i}{4\sqrt{6}} h_{\tI}  e^a
\left( \g_{abc}\ve_i - 4 \eta_{ab} \g_c \ve_i\right)
\widehat{\cH}^{bc\,\tI} - \delta_\ve q^X {\omega_{X\,i}}^j \psi_j
+   \nonumber \\
&-& \frac{1}{12} e^a \g_{ab} \ve^j \;
\bl^{\tc}_i \g^b \l_j^{\tc} + \frac{1}{48} e^a \g_{abc}
\ve^j \;  \bl_i^{\ta} \g^{bc} \l_j^{\ta} +
\frac{1}{6} e^a \ve^j \;  \bl_i^{\tc} \g_a \l_j^{\tc}
+ \nonumber \\
&-& \frac{1}{12} e^a \g^b\ve^j \;
\bl_i^{\tc} \g_{ab} \l_j^{\tc}  + \frac{1}{8} e^a
\g^{bc} \ve_i \; \bz_A \g_{abc} \z^A
+\frac{i}{\sqrt{6}} g_R \, e^a \g_a \ve^j P_{ij}, \\
\delta_{\ve}
\phi^{\tx} &=&  \frac{i}{2} \bar{\ve}^i \l_i^{\ta} \,
f_{\ta}^{\tx}, \\
\delta_{\ve} A^I &=& \th^I, \hbox{ where }
\th^{\tI} \equiv - \frac{1}{2} e^a \,  \bar{\ve}^i \g_a
\l_i^{\ta}\, h_{\ta}^{\tI} + \frac{i\sqrt{6}}{4}
\ol{\psi}^i\ve_i \, h^{\tI},\\
\delta_{\ve} \l_i^{\ta} &=& -
\frac{i}{2} f^{\ta}_{\tx} \g^a \ve_i \, \widehat{\cD}_a \phi^{\tx}
- \delta_\ve \phi^{\tx} {\Omega_{\tx}}^{\ta\tb} \l^{\tb}_i -
\delta_\ve q^X \omega_{X\,i}{}^j \l^{\ta}_j +\frac{1}{4}
h_{\tI}^{\ta} \g^{ab} \ve_i \,  \widehat{\cH}_{ab}^{\tI}  +
\nonumber \\
&-& \frac{i}{4\sqrt{6}} {T^{\ta}}_{\tb\tc} \left[ -3
\ve^j \, \bl^{\tb}_i \l_j^{\tc} + \g_a\ve^j \,
\bl^{\tb}_i \g^a \l_j^{\tc} + \frac{1}{2} \g_{ab}
\ve^j  \, \bl^{\tb}_i \g^{ab} \l_j^{\tc} \right]+ \\
&+& g_R \ve^j P^{\ta}_{ij} + g  W^{\ta} \ve_i,  \nonumber \\
\delta_{\ve} B^M &=& d\th^M + \frac{i}{8} g  e^b e^a \,
\bar{\ve}^i \g_{ab} \l_i^{\ta} \, h_{N} \Omega^{MN} +
g \frac{\sqrt{6}}{8} e^a \, \ol{\psi}^i \g_a \ve_i \,
\Omega^{MN} h_N, \\
\delta_{\ve} q^X &=& - i \bar{\ve}^i \z^A \, f_{iA}^X, \\
\delta_\ve \z^A &=& - \frac{i}{2}\g^a \ve^i \widehat{\cD}_a q^X \,
f_{iX}^A - \delta_\ve q^X {\omega_{X\,B}}^A \z^B +  g  \ve^i \cN_{i}^A.
\eea
The hatted quantities $\widehat{\phantom{ab}}$ are the
supercovariantization of the unhatted ones.

\section{The action}

\def\he{\hat{e}}

We now turn to the Lagrangean of the $\cN = 2$, $D = 5$
gauged supergravity in interaction with vector, tensor and hypermultiplets.
We take as a start the results of \cite{MaxEin,gaugetens}, and add all the
modifications that are needed in presence of hypermultiplets and
$SU(2)_R$ gauging.
We'll exhibit our result at the {\it component level},
and for brevity, whenever we use the same symbols as in section 3,
we now mean those objects evaluated at $\th = 0 = d \th$.
In particular, the superspace differential $d$ now becomes the ordinary
differential.

Each form can be decomposed along the vielbeins $e^a = dx^\mu
{e_\mu}^a$ and the gravitino, which reduces to $\psi^i = dx^\mu \psi_\mu^i
\equiv e^a \psi_a^i$.
The supercovariant connection one--form $\hat\omega_a{}^b = dx^\mu
\hat\omega_{\mu\,a}{}^b$ is naturally introduced, via equation
\eqn{fondvin}, now evaluated at $\th = 0 = d \th$, as
\beq
\label{om}
de^a + e^b \,\hat\omega_b{}^a = -\frac{1}{4} \ol\psi^i \g^a \psi_i.
\eeq
This determines $\omega$ as the metric connection, augmented by the
standard gravitino bilinears.
The supercovariant curvature two--form becomes ${R_a}^b = d
\hat\omega_a{}^b + \hat\omega_a{}^c \hat\omega_c{}^b$, 
with $\hat\omega$ given in \eqn{om}.

Since we write the Lagrangean as a five--form, it is also convenient
to define the $(5-p)$--forms
\beq
\he^{a_1 \ldots a_p} \equiv - \frac{1}{(5-p)!} \e^{a_1 \ldots a_p b_1
\ldots
b_{5-p}} e_{b_1} \ldots e_{b_{5-p}} .
\eeq
In particular $\he = \sqrt{-g} \,d^5x$.

Invariance of the action under supersymmetry can be checked by
using the standard trick of lifting this action to superspace, performing the
superspace differential and taking the interior product with $\ve^A$.
It is obvious that the equation \eqn{28} gives
$$
\delta_{\ve} \int \cL = \int i_\ve \cD \cL + \int \cD i_\ve \cL
$$
and $\cD i_\ve \cL$ is a total derivative  that we can discard.

\medskip

The Lagrangean for the gauged theory can be split  as
$$
\cL = \cL_{KIN} + \cL_{Pauli} + \cL_{CS} + \cL_{mass} + \cL_{pot}
+\cL_{4Fermi},
$$
where $\cL_{KIN}$ contains the kinetic terms, $\cL_{Pauli}$ describes
the couplings between the bosonic field--strengths and the fermions,
$\cL_{CS}$ contains the Yang--Mills Chern--Simons term, $\cL_{mass}$
gives the mass of the fermions, $\cL_{pot}$ contains the typical
potential of the gauged theories and $\cL_{4Fermi}$ contains the
four--Fermi terms.

The kinetic terms are
\bea
\cL_{KIN} &=& \frac{1}{2} R^{ab}
\he_{ab} + \frac{i}{4} e^a e^b  \ol{\psi}^i
\g_{ab} \cD \psi_i -\frac{1}{2} \bl^{i\ta} \g_a \cD \l_i^{\ta} \he^a
-
\bz^A \g^a \cD\z_A \he_a + \nonumber \\
&+& \left[ \frac{1}{2} g_{\tx\ty} Q^{\tx}_a Q^{a\ty}
\he - \he^a g_{\tx\ty} Q_a^{\tx} \left(\cD \phi^{\ty} - \frac{i}{2}
\ol{\psi}^i \l_i^{\ta} f_{\ta}^{\ty}\right) \right] + \nonumber \\
&+& \left[ \frac{1}{2} g_{XY} Q^{X}_a Q^{a Y}
\he - \he^a g_{XY} Q_a^{X} \left(\cD q^Y + i \ol{\psi}^i
\z^A f_{iA}^Y \right) \right] + \\
&+& \frac{1}{4} a_{\tI\tJ} Q^{\tI}_{ab} \left[
Q^{\tJ\,ab}
\he + 2 \he^{ab} \left(\cH^{\tJ} + \frac{1}{2} e^c \,\ol{\psi}^i \g_c
\l_i^{\ta} \, h_{\ta}^{\tJ} -\frac{i\sqrt{6}}{8} \ol{\psi}^i
\psi_i h^{\tJ} \right) \right] + \nonumber \\
&+& \frac{1}{g } \Omega_{MN} B^M H^N,  \nonumber
\eea
where $Q^{\tx}_a$, $Q_a^{A}$ and $Q^{\tI}_{ab}$ are auxiliary
fields which have been introduced in order to write the Lagrangean as
a five--form.
Their equations of motion give
\bea
e^a Q^{\tx}_a &=& \cD \phi^{\tx} - \frac{i}{2} \ol{\psi}^i \l_i^{\ta}
f_{\ta}^{\tx}, \nonumber \\
e^a Q_a^{X} &=& \cD q^X + i \ol{\psi}^i \z^A f_{iA}^X, \\
\frac{1}{2} e^a e^b Q^{\tI}_{ba} &=& \cH^I + \frac{1}{2} e^c
\, \ol{\psi}^i \g_c \l_i^{\ta} \, h_{\ta}^{\tI}
-\frac{i\sqrt{6}}{8}
\ol{\psi}^i \psi_i h^{\tI}, \nonumber
\eea
which, upon substituting this back in $\cL_{Kin}$, yields the
usual super--covariantized kinetic terms.

The Pauli--like couplings are described by:
\bea
\cL_{Pauli} &=& \frac{i}{4} \, \bl^{i\ta} \g_{ab} \psi_i\,
\cH^{\tI}h_{\tI}^{\ta} \; e^a e^b - \frac{\sqrt{6}}{8} \,
\ol{\psi}^i \g_a \psi_i\,\cH^{\tI} h_{\tI} e^a - \frac{i}{4}
\bl^i_{\ta} \g_{ab} \psi_i \, \cD\phi^{\tx} f_{\tx}^{\ta} \he^{ab} +
\nonumber \\
&-& \frac{i}{4} \Phi_{\tI\ta\tb} \,\bl^{i\ta} \g_{ab} \l_i^{\tb}
\,\cH^{\tI} \he^{ab} + \frac{i}{4\sqrt{6}} \,\z_A \g_{ab}
\z^A \,\cH^{\tI} h_{\tI} \he^{ab} +\\
&+&\frac{i}{2} \,\ol{\psi}^i \g_{ab}
\z^A \, f_{Ai}^X \cD q^Y g_{XY} \he^{ab} , \nonumber
\eea
where
$$
\Phi_{\tI\ta\tb} \equiv \sqrt{\frac{2}{3}}\left(
\frac{1}{4}\delta_{\ta\tb}
h_{\tI} + T_{\ta\tb\tc} h^{\tc}_{\tI}\right),
$$
whereas the Chern--Simons couplings are fixed as usual to
\beq
\cL_{CS} = \frac{2}{3\sqrt{6}} C_{IJK} \left( F^I F^J A^K -
\frac{3}{4} g  F^I A^J A^L A^F f^K_{LF}  + \frac{3}{20} g^2 f^J_{GH}
f^K_{LF} A^I A^G A^H A^L A^F \right).
\eeq

The four Fermi terms can further be split, following
$$
\cL_{4Fermi} = \cL_{4\l} + \cL_{3\l\psi} +\cL_{other},
$$
such that
\bea
\cL_{4\l} &=& \he \left[ \frac{1}{48\sqrt{6}} \,
\bl^{i\ta} \g_{ab} \l_i^{\tb}\;
\bl^{j\tc} \g^{ab} \l_j^{\td}\, T_{\ta\tb\tc;\td} +
\frac{1}{24} K_{\ta\tb\tc\td} \left( 2 \,
\bl^{i\ta} \l_i^{\tb}\;\bl^{j\tc}
\l_j^{\td}\, + \right.\right. \nonumber \\
&+& \left. \left. \,\bl^{i\ta} \g_a\l_i^{\tb}\;
\bl^{j\tc} \g^a \l_j^{\td}\,\right)- \frac{1}{12} \,
\bl^{i\ta} \l^{j\ta}\;\bl_j^{\tb}
\l_i^{\tb}\, + \right. \nonumber \\
&-& \left. \frac{1}{24} \,\bl^{i\ta} \g_a\l^{j\ta}\;\bl_j^{\tb}
\g^a \l_i^{\tb}\,+\frac{1}{64} \,\bl^{i\ta}
\g_{ab}\l^{j\ta}\;
\bl_j^{\tb} \g^{ab} \l_i^{\tb}\,\right]
\eea
and
\beq
\cL_{3\l\psi} = \frac{2i}{3\sqrt{6}} T_{\ta\tb\tc} \left[
\bl^{i\ta} \psi^j\;\bl_i^{\tb} \g_a \l_j^{\tc}- \frac{1}{2}
\, \bl^{i\ta} \g_a \psi^j\; \bl_i^{\tb}
\l_j^{\tc} \right] \he^a
\eeq
exactly reproduce the terms of the Einstein--Maxwell theory presented
in \cite{MaxEin}.

$\cL_{other}$ contains  the new four--Fermi interactions with
the hypermultiplet spinors and reads:
\bea
\cL_{other} &=& -\frac{i}{32} e^a \, \ol\psi^i \g_a \psi_i\,
\ol\psi^j \psi_j +  \he_{ab} \left[ \frac{1}{32} \,\ol\psi^i \psi_i\,
\bz_A \g^{ab} \z^A + \frac{1}{4} \,\bz_A \g^{ab} \psi_i \,
\ol{\psi}^i \z^A \right]+ \nonumber \\
&+& \left[ \frac{1}{16} \, \ol\l^{\ta i} \g_{ab}
\psi_i\,
\bl^{\ta k} \psi_k + \frac{1}{16} \, \ol\l^{\ta i} \g^{c}
\psi_i \; \bl^{\ta j} \g_{abc} \psi_j\,
- \frac{1}{64} \,\bl^{\ta i}
\g_{ab} \l_{\ta i} \, \ol\psi^k \psi_k \right]
\he^{ab}+\nonumber \\
&+& \he \frac{1}{16} \, \bz_A \g^{ab} \z^A
\; \bz_B \g_{ab}\z^B - \frac{1}{4} \Omega_{ABCD}
\left( 5 \bz^A \z^B \; \bz^C \z^D - \bz^A \g^a \z^B \; \bz^C \g_a
\z^D\right)
\nonumber \\
&+& \he \frac{1}{16} \, \bz_A \g^{abc}
\z^A \; \bl^{\ta i}  \g_{abc} \l_i^{\ta}.
\eea

We point out that there are no terms with three gravitinos and
one gluino or two $\z$, one gluino and one gravitino due to the
orthogonality properties of $h^{\tI}$ and $h^{\tI\ta}$.

We can finally describe the mass and potential terms.
The first reads
\bea
\cL_{mass} &=& g_{R} \left[ \frac{i}{2\sqrt{6}} \,
\bl^{i\ta} \l^{j\tb} \,P^{\ta\tb}_{ij} \, \he-\, \bl^{i\ta}
\g_a \psi^j\, P^{\ta}_{ij} \he^a - \frac{i\sqrt{6}}{8} \,
\ol\psi^i \g_{ab} \psi^j \, P_{ij} \he^{ab}\right] +
\nonumber\\
&+& g  \left[ \, \bl^{i\ta} \g_a \psi_i \, W^{\ta}\,  \he^a
+ \bl^{i\ta} \l^{\tb}_j\,W_{\ta\tb} \, \he +
2 \, \ol\psi^i \g_a \z^A \, \cN_{iA} \, \he^a
+\right. \nonumber\\
&-& \left. 2 i \, \bz^A \l^{\ta i}\, \cM_{Ai\ta}\,  \he +
 \bz^A \z^B \; \cM_{AB} \, \he\,  \right] ,
\eea
where $P^{\ta}_{ij}$, $P_{ij}$, $W^{\ta}$ and $\cN_{iA}$ were defined
in the previous section by equations \eqn{Paij}, \eqn{Pij}, \eqn{Wa} and \eqn{NiA},
and the other mass matrices are defined as
\bea
P^{\ta\tb}_{ij} &\equiv& \delta^{\ta\tb} P_{ij} + 4 T^{\ta\tb\tc}
P^{\tc}_{ij},\\
W^{\ta\tb} &\equiv& i h^{I [\ta}K^{\tb]}_{I} + \frac{i\sqrt{6}}{4}
h^I  K^{\ta;\tb}_I ,\\
\cM_{Ai\ta} &\equiv& f_{AiX} K^X_I h^{I\ta},\\
\cM_{AB} &\equiv&  \dfrac{i\sqrt{6}}{2} f_{AX}^i f_{BiY}
K^{[Y;X]}_I h^I.
\eea
This mass term is of first order in the gauge coupling constants and has
coefficients fixed by variations of the kinetic terms.

Finally, the potential is ($\cL_{Pot} = - \cV \he$):
\beq
\cV = 2 g^2 W^{\ta} W^{\ta} - g^2_R \left[ 2
P_{ij} P^{ij} -P^{\ta}_{ij} P^{\ta\,ij}\right] + 2 g^2 \cN_{iA} \cN^{iA}.
\eeq

As an outcome of this complete analysis, one can remark many
similarities with the analogous four--dimensional matter coupled
theory \cite{D4N2}.
However, a first difference is the existence of
tensor multiplets satisfying a first order equation of motion,
and of a corresponding new term in the scalar potential.
The second lies in the presence of the Chern--Simons term, that as well known
is a peculiar feature of odd-dimensional space--times.
Moreover, the geometry described by the scalars of vector multiplets
is now ``very special'' rather than  special K\"ahler,
and thus the $U(1)$ K\"ahler connection does not exist and
all the 4D structures deriving from its gauging are missing. Finally,
as already remarked, the Yang--Mills gauging in $D=5$
does not give
rise to any contribution to the scalar potential unless
some of the ungauged vectors are dualised into tensor multiplets.

\section{Some comments on the scalar potential}

We have found that the bosonic sector of $5D$, $\cN=2$
supergravity is described by the Lagrangean:
\bea
\he^{-1}\cL^{\cN = 2}_{bosonic} &=& - \frac{1}{2} R -
\dfrac{1}{4} a_{\tI\tJ} \cH^{\tI}_{\mu\nu} \cH^{\tJ \mu\nu} -
\dfrac{1}{2} g_{XY} \cD_\mu q^X \cD^\mu q^Y + \nonumber \\
&-& \dfrac{1}{2} g_{\tx\ty} \cD_\mu \phi^{\tx} \cD^\mu \phi^{\ty} +
\frac{\he^{-1}}{6\sqrt{6}} C_{IJK} \e^{\mu\nu\rho\s\tau}
F_{\mu\nu}^I F_{\rho\s}^J A^K_\tau + \\
&+& \frac{\he^{-1}}{4g} \e^{\mu\nu\rho\s\tau} \Omega_{MN}
B^M_{\mu\nu} \cD_\rho B^N_{\s\tau} - \cV(\phi,q), \nonumber
\eea
where
\beq
\label{potential}
\cV = 2 g^2 W^{\ta} W^{\ta} - g^2_R \left[ 2
P_{ij} P^{ij} -P^{\ta}_{ij} P^{\ta\,ij}\right] + 2 g^2 \cN_{iA} \cN^{iA}.
\eeq
The bosonic part of the supersymmetry transformation rules is
given by
\bea
\label{susygrav}
\delta_{\ve} \psi_{\mu i} &=&
\cD_\mu(\widehat{\omega}) \ve_i + \frac{i}{4\sqrt{6}} h_{\tI}
\left( \g_{\mu\nu\rho}\ve_i - 4 g_{\mu\nu} \g_\rho \ve_i\right)
\widehat{\cH}^{\nu\rho\,\tI}
+\frac{i}{\sqrt{6}} g_R \,  \g_\mu \ve^j P_{ij}, \\
\label{susylam}
\delta_{\ve} \l_i^{\ta} &=& -
\frac{i}{2} f^{\ta}_{\tx} \g^\mu \ve_i \, \widehat{\cD}_\mu \phi^{\tx}
 +\frac{1}{4}
h_{\tI}^{\ta} \g^{\mu\nu} \ve_i \,  \widehat{\cH}_{\mu\nu}^{\tI}
+ g_R \ve^j P^{\ta}_{ij} + g  W^{\ta} \ve_i,   \\
\label{susyzet}
\delta_\ve \z^A &=& - \frac{i}{2}f_{iX}^A \g^\mu \ve^i \widehat{\cD}_\mu q^X
 +  g  \ve^i \cN_{i}^A,
\eea
where
\bea
\label{pij}
P_{ij}& \equiv& h^I P_{I\, ij},\\
\label{paij}
P^{\ta}_{ij} &\equiv& h^{\ta I} P_{I\, ij}, \\
W^{\ta} &\equiv& -\frac{\sqrt{6}}{8} \Omega^{MN} h_M^{\ta} h_N =
 \frac{\sqrt{6}}{4} h^I K_I^{\tx} f_{\tx}^{\ta},\\
\cN^{iA} &\equiv& \frac{\sqrt{6}}{4} h^{I} K^X_{I} f_{X}^{Ai}.
\eea

As expected,  one sees that the scalar potential of the gauged
supergravity theory is
constructed out of the squares of the fermion shifts that arise
in the supersymmetry
transformations due to the gauging.

The four terms in the potential \eqn{potential} have different origins.
Those of order $g^{2}_R$ come from the
 $R$--symmetry gauging, whereas those of order
$g^2$ come from the gauging of the Yang--Mills group $K$.
In detail, the YM ones are given by the squares of the $g $ order
shifts in the supersymmetry transformation laws of the $\l_{i}^{\ta}$
and $\z^{A}$ fields, while the $R$--symmetry ones are given by the
square of the order $g_R$ shifts in the $\psi^i$ and $\l_{i}^{\ta}$
supersymmetry variations.
As in the four--dimensional case
\cite{D4N2}, this can be related to the existence of a  ``Ward
identity" for the scalar potential \cite{old}, but here we don't need to use the
\eqn{prep2} condition on the prepotential to ensure it, due to the
reality properties of the very special manifold $\cS$ parametrised by the
 $h^I$ coordinates.

In presence of hypermultiplets, one 
  generically has in the fermionic shifts the
$SU(2)_R$--valued quantities $P_{ij}(\phi,q)$
and $P^{\ta}_{ij}(\phi,q)$ of \eqn{pij} and \eqn{paij}
containing the prepotential
\beq
P_{Iij}\equiv i P_I^r(q) (\sigma_r)_{ij}\ \ \ r=1,2,3\ ,
\eeq
where $(\s_r)_i{}^j$ are the usual Pauli matrices,
in place of the $P_0 \delta_{ij}$ and
$P^{\ta}\delta_{ij}$ of \cite{U1g,gaugetens,potenz}.

\medskip

For our metric signature, this potential allows for the existence of
Anti de Sitter vacua if  $\cV(q^*, \phi^*) < 0$ for $\cV^{\prime} (q^*,
\phi^*) = 0$. Thus it
 is  straightforward to see that the only contribution
which can allow for such solutions is the $2 P_{ij} P^{ij}$ term,
coming from the $R$--symmetry gauging of the gravitinos.
This implies that a simple Yang--Mills gauging, even in presence
of both tensor and hypermultiplets, does not allow Anti de Sitter
solutions.

Since \eqn{potential}  is the most general potential of $5D$,
$\cN = 2$ gauged supergravity coupled to all matter,
all the previously studied examples must be found as peculiar subcases.

We first analyze the choice $n_H = 0$.
Since in this case the $\cQ$--manifold disappears, we are forced to
put $K^X_I = 0$, and the Yang--Mills sector is reduced
to the $W^2$ part due to the tensor multiplets.
Moreover, the absence of the quaternionic fields $q^X$ implies that
the prepotentials are set to zero, or at most are $SU(2)$--valued
constants.
Their most general form is now given by
\beq
P_{I\,i}{}^j = i \xi_I^r \, (\s_r)_i{}^j,
\eeq
where $\xi_I^r$ are three real constants generically\footnote{
In some special cases \cite{GZ}, one can still preserve the
full $SU_R(2)$ gauging by the choice
$\xi^r_I = \delta_I^r$ and thus identifying by \eqn{52} the $SU_R(2)$
structure constants with those of a $SU(2) \subset K$:
$g f^r_{st}=g_R {\e^r}_{st}$.} breaking $SU(2) \to U(1)$.

The \eqn{prep2} condition  becomes
\beq
\label{52}
g \, f^K_{IJ} \, \xi_K^r = g_R \, \e^{rst} \, \xi_I^s \, \xi_J^t.
\eeq
If one makes the choice $\xi_I = (0, V_I, 0)$, the condition \eqn{52}
reduces to
\beq
f^K_{IJ} V_K = 0,
\eeq
which is the supersymmetry requirement of \cite{gaugetens}.
In particular, all the results therein
can be recovered by substituting
\beq
P_{I\,ij} = V_I \delta_{ij}
\eeq
in the action and supersymmetry laws.
The above mechanism is the local analogue of the Fayet--Iliopoulos
phenomenon occurring also in four dimensions \cite{D4N2}.

The $n_T=0$ case is trivial, as it simply removes the $g^2 W^2$
term.
This leaves us with  the potential presented in \cite{hyper+sugra},
where the gauge group was chosen as $K=U(1)^{n_V+ 1}$ and $g_R = g$.

More interesting is to take  $n_V = 0$.
This implies that there is only one vector: the graviphoton.
The potential is still non--vanishing, and becomes
\beq
\cV = \frac{3}{4} g^2 K^X K_X - 2 g_R^2 P_{ij} P^{ij},
\eeq
which in principle could admit Anti de Sitter vacua.

The $n_V = n_T = n_H = 0$ case gives back the pure gauged
supergravity
of \cite{gaugedpura}, with the potential
\beq
\cV = - 4 V^2,
\eeq
for the choice $P_{ij} = V \delta_{ij}$ ($\xi = (0, V, 0)$).
This is, of course, the Anti de Sitter five--dimensional supergravity.

\bigskip

It is now relevant to consider the possible existence of smooth Randall--Sundrum
domain--wall solutions of type RS2 providing an ``alternative to compactification".
An easy way to do it is to determine
the vacua obtained from the full potential \eqn{potential}  and check whether
they are of the same nature as
those already studied in \cite{U1g,KL,potenz}.

It has been shown \cite{KL} that in absence of tensor multiplets
there are no RS2 solutions at all, while if tensor multiplets are
added one still excludes the presence of supersymmetric solutions
but leaves open the possibility of having non--BPS ones.

Although we have not yet performed a complete analysis of \eqn{potential},
we can already make some comments on the supersymmetric
vacua  and BPS solutions of the full theory.

As for the theory without hypermultiplets
\cite{KL,potenz}, the cosmological constant of an $\cN = 2$
supersymmetric vacuum  is only given by
\beq
\label{cosmo}
\cV(\phi^*,q^*)=g_R^2 \, P_{ij} P^{ij}
(\phi^*,q^*).
\eeq
Using the orthogonality property
$$
W^{\ta} P^{\ta}_{ij} = 0,
$$
one can easily show that the requirement $\langle \,\delta_\ve \l\,\rangle =
\langle\,\delta_\ve \z\,\rangle =
0$ implies that an $\cN = 2$ supersymmetric ground state must
satisfy
\beq
\label{punto}
\langle\, W^{\ta}\,\rangle = \langle\,P^{\ta}_{ij}\,\rangle =
\langle\,\cN_{iA}\,\rangle = 0.
\eeq
Therefore, the only non--trivial effect of the
$W^{\ta}$ and $\cN_{iA}$ terms can be a change in
the shape of the critical point.

The above result can also be obtained  by a different argument\footnote{We are
indebted to R. Kallosh for explaining this argument
and for important discussions on the results below.}.
The integrability condition on the gravitino supersymmetry rule
\eqn{susygrav}
imposes that the vacuum expectation value is given by \eqn{cosmo}.
Thus, in order to have  non vanishing $\langle\, W^{\ta}\,\rangle$,
$\langle\,P^{\ta}_{ij}\,\rangle $ and
$\langle\, \cN_{iA}\,\rangle$ they must compensate each other in the
potential.
However, being all positive squares, they can never cancel
unless they vanish.
This  means that the $\cN = 2$ supersymmetric critical
points of the full $\cN = 2$, $D=5$ gauged supergravity theory
have the same nature as those of the reduced theory analyzed in
\cite{KL}.

There remains to examine the possibility of having
lower supersymmetric BPS solutions.

To do this, one tries to relax the  conditions \eqn{punto} near the
critical point, where scalars are not fixed anymore.
It has been verified that in absence of tensor and hyper--multiplets
the derivative of scalars in the radial direction $y$
can be chosen proportional to the derivative of the
superpotential, leading to solutions that
preserve half supersymmetry.
When tensor multiplets are added, it turns out \cite{KL} that
relaxing the condition $W^{\ta} =0$,  forbids
any supersymmetric solution.

We now show along the lines of \cite{KL} that the same
phenomenon occurs also in the most general case
where hyper--multiplets are added
and the full $SU(2)_R$ group is gauged.

We choose the $y$ dependence of all fields appearing in the
gluino susy rule \eqn{susylam} as
$$
\partial_y\phi^{\ta}(y)\sim W^{\ta}(\phi)\sim
P^{\ta}_{ij}(\phi,q).
$$
For BPS solutions to exist, one has to
find some Killing spinors which, in addition to the usual
constraint coming from the gravitino susy rule \eqn{susygrav}
\beq
\label{cond1}
i \g^y \ve_{i} \sim g_R P_{ij} \ve^j,
\eeq
also satisfy
\beq
\label{cond2}
\delta_\ve \l_i^{\ta}={{\cA^{\ta}}_i}^j \ve_j = 0
\eeq
where the operator matrix $\cA^{\ta}$ can be read off from
\eqn{susylam}. This amounts to finding a Killing spinor
eigenvector of  $\cA^{\ta}$ with zero eigenvalue, i.e. the
matrix $\cA^{\ta}$ must be degenerate.
It can be seen that even if with the full R--symmetry
gauging the $P_{ij}$ and $P^{\ta}_{ij}$ are $SU_R(2)$--valued matrices,
this condition still has no solutions.
Indeed, upon substituting \eqn{cond1} into the gluino
transformation rule \eqn{susylam}, and
requiring that all bosonic functions of the scalars have the same
$y$
behaviour, the $\cA^{\ta}$ operator reduces to the form
\beq
{{\cA^{\ta}}_i}^j \sim i Q^{r\ta} (\s_r)_i{}^j + W^{\ta}
\delta_i^j,
\eeq
where  $Q^{r\ta}$ indicates some real combination of the various
$y$--dependent quantities
\beq
Q^{r\ta}(y)\sim\partial_y \phi^{\ta}P^r+P^{r\ta},
\eeq
and the obvious notation $P_{ij}\equiv i P^r (\sigma_r)_{ij}$,
$P^{\ta}_{ij}\equiv i P^{r\ta} (\sigma_r)_{ij}$.
Requiring  det$\cA^{\ta} = 0$ (or equivalently for a projector
$\cA^2=\cA$), imposes for each ${\ta}$
\beq
(W^{\ta})^2 + (Q^{\ta})^2 = 0,
\eeq
which has no solutions except for $W^{\ta} = 0 = Q^{\ta}$ and this
takes us back to the cases of \cite{KL}.

This seems to rule out the presence of BPS
solutions, at least when $h_{\tI} \cH^{\tI} = h^{\ta}_{\tI} \cH^{\tI} =
0$.
The only open possibilities appear to be either BPS solutions with
non--trivial electric or magnetic fields $\cH^{\tI}$, or non--BPS solutions.
It must be noted, however, that also the examples considered up to
now with $\cH^{\tI} \neq 0$ do not seem to admit solutions
of the RS2 type \cite{BCS}, i.e. with a metric of the form
$$
ds^2 = a(y)\, dx^2 + dy^2,
$$
approaching asymptotically $AdS_5$.

We leave the investigation of all these more general cases to
future work.

\section{The low--energy theory for the $AdS_{5} \times T^{11}$
compactification}

We collect here, as a further application of the matter coupled
5D gauged supergravity,
some comments about the structure of the theory corresponding to the
type IIB compactification on $AdS_5 \times T^{11}$ \cite{T11b}.
This study could reveal very useful for further analyzing
the $AdS/CFT$ correspondence in such  non--trivial case.

One is interested in the construction of the low--energy
theory in order to study the deformations and RG fluxes of the dual field theory.
Once a specific scalar manifold $\cM$  is chosen,
one can study the stationary points of the potential and look for
solutions interpolating between them.
This should correspond to a flux in the CFT side.

As pointed out by S. Ferrara, a very interesting deformation of the
four--dimensional conformal field theory \cite{T11a} is given by the
operator
\beq
Tr\left( A_i B_j A_k B_l\right) (\s^r)^{ik} (\s^s)^{jl} \delta_{rs},
\eeq
which has conformal dimension $\Delta = 3$ and therefore corresponds
to a marginal deformation preserving supersymmetry.

The generic superpotential  now is given by
\beq
\cW = \left[ \l \, \e^{ij} \e^{kl} + \mu \, (\s^r)^{ik}
(\s_r)^{jl}\right] \; Tr\left( A_i B_j A_k B_l\right),
\eeq
where $\l$ and $\mu$ are two coupling constants
and the isometry group has now been broken from $SU(2) \times SU(2)$
to the diagonal $SU(2)$.

From the supergravity point of view, this implies that there should
exist
another $\cN = 2$ vacuum of five--dimensional supergravity with this
symmetry group.
Lifting this solution to ten dimensions should give a
metric which is the warped product of  $AdS_5$ and a deformation of $T^{11}$
with isometry reduced to $SU(2)_{diag}$.

An important point is that
the $T^{11}$ compactification does not seem to be a
solution  of the $\cN =8$ theory
\cite{D5N8}.
If such solution would exist, we should find a stable vacuum with
$G = SU(2)\times SU(2)$ isometry preserving $\cN = 2$ supersymmetry.
There are two possible embeddings of $G$ in the $\cN = 8$ gauge group $SO(6)
\simeq SU(4)$, leading to two families of vacua:
$$
i) \quad 6 \to (2,2) + (1,1)+ (1,1),
$$

$$
ii) \quad 6 \to (3,1) + (1,3).
$$

The first family can be obtained by turning on simultaneously the two
scalars $\l$ and $\mu$,
which break $SO(6) \to SO(4) \times SO(2)$ and $SO(6) \to SO(5)$.
According to the analysis of  \cite{D5N8},
this gives a supersymmetry operator  of the form
\beq
W_{ab} =-\frac{1}{4} \left( 4 e^{\l+\mu} + e^{\mu-\l} +
e^{-\l-\mu}\right)\delta_{ab},
\eeq
in the gravitinos supersymmetry transformations,
which cannot preserve $\cN = 2$.
In fact, in order to preserve  $\cN$ supersymmetries, the operator $W_{ab}$
must have, at the critical  point, $\cN$ eigenvalues
of the form
\beq
\label{crit}
\pm \sqrt{-\frac{3}{g^2}\cV|_*}.
\eeq

To obtain vacua of the type (ii),
we must study the potential given by the
three scalars respecting this invariance \cite{D5N8}.
We find that the potential becomes in this case
\beq
\cV = - \frac{3}{8} g^2 \left[ 3 f^2(\l) ( \a^2 + \b^2) \sinh^2 \l + 3
\cosh^2 (2\l) - \cosh(4\l)\right],
\eeq
where $\l$, $\a$, $\b$ are the three scalars and $f(\l)\equiv \frac{e^\l -
e^{-2\l}}{3\l}$.
At the extremum  one finds the potential $\cV = - \frac{3}{4}g^2$ and
the supersymmetry operator $W_{ab} = -\frac{3}{2}
\delta_{ab}$.
Using \eqn{crit} one finds that eight supersymmetries are
preserved, and one retrieves the  highest symmetric $S^5$ solution.

In conclusion, the aforementioned flux from the $T^{11}$ to a solution
with residual $SU(2)_{diag}$ symmetry can only be studied within
the  $\cN = 2$, $D = 5$ gauged theory corresponding to the $T^{11}$
low--energy model.

To build this model, we need all the predictive power of the $AdS/CFT$
correspondence and the spectrum analysis performed in \cite{T11b}.
As for the maximally symmetric cases of type IIB on $AdS_5 \times S^5$ or
$M$--theory on $AdS_{4/7} \times S^{7/4}$, we expect that the low--energy
gauged supergravity states correspond to the theory given by fields which are
the products of two singletons.
This means that their masses are less than or at most equal to zero.

Using the \cite{T11b} results and notations, the theory should be described by
the massless graviton multiplet, corresponding to the stress--energy
tensor $W_\a \bar{W}_{\dot{\a}} + \ldots$ from the boundary point of view,
the seven massless vector multiplets corresponding to the $SU_A(2) \times
SU_B(2) \times U_{Betti}(1)$ conserved currents $A_i\bar{A}_j$, $B_i\bar{B}_j$ and
$Tr(A_i\bar{A}^i + B_i\bar{B}^i$), and six hypermultiplets
corresponding to the $Tr(AB)$ and $W^2$ operators.

There are no tensor multiplets, since they should be described by the product
of at least three singleton states $Tr [W_\a (AB)]$.

Although this is not enough to uniquely fix the scalar
manifold $\cM = \cS \otimes \cQ$, we can still  extract some useful information.
For example, the $\cQ$ manifold must contain at least the $G= SU(2) \times
SU(2)$ isometries and two zero--modes: i.e. two of its scalars must be
massless, since they have to correspond to the moduli of the conformal field
theory.
The correct manifold must anyhow fall in the classification of \cite{dWvP}.

We can say something more on the $\cS$ manifold of the vector multiplet
scalars.
As already claimed in \cite{T11b}, the $AdS/CFT$ foresees the value of the
Chern--Simons couplings $C_{\tI\tJ\tK}$ through the computation of the
anomalies in the boundary theory.
The result is that the polynomial \eqn{poly} describing the $\cS$ target
manifold is given by
\beq
\label{63}
\a \, \xi_r^3 + \b\, \xi_r(\xi_A^2 + \xi_B^2) + \g \, \xi_r \xi_b^2 + \delta
\, \xi_b(\xi_A^2 - \xi_B^2) = 1,
\eeq
where $\a, \b, \g$ and $\delta$ are constants and  $\xi_r$,
$\xi_b$, $\xi_A$ and $\xi_B$ denote the ambient coordinates corresponding to the
$R$--symmetry, Betti and $SU_A(2) \times SU_B(2)$ symmetries.

It is known that there always exists a point $c^I$ where the metric becomes
flat ($\delta_{IJ}$) and the cubic polynomial takes the standard form.
In our case this happens for $c_r = 1$ and $c_A = c_B = c_b = 0$ and fixes
\beq
\label{64}
\a = 1, \quad \b = \g = -\frac{1}{2},
\eeq
whereas $\delta$ is still free.

At this point one can go farther and see whether this manifold corresponds to one
of the homogeneous spaces classified in \cite{dWvP}.
Unfortunately, it is easy to prove that the given polynomial cannot correspond
to a homogeneous space, as there is no $SO(7)$ rotation reducing the $\delta$
piece of \eqn{63} to the form of one of the three
families classified in \cite{dWvP}.

Anyway, this does not spoil our hope to study the minima of the potential for
such  a model in the future, since the cubic surface is specified by \eqn{63}
and \eqn{64} up to the $\delta$ coefficient, which can be computed
explicitly evaluating the three--point functions of the corresponding
anomalies.

\vskip 1truecm

\paragraph{Acknowledgements.}
\ We are glad to thank S. Ferrara for helpful conversations and A. van Proeyen
also for pointing out some misprints.
We have also profited greatly from discussions with R. Kallosh and we thank
her for clarifying to us the peculiarities of the various brane--world scenarios.
We are particularly grateful to R. D'Auria for his interest in our work,
the numerous valuable comments and for his continuous encouragement. \\
This research is supported in part by EEC under TMR
contract ERBFMRX-CT96-0045.

\vskip 1truecm

\section*{Appendix A: Notations and conventions}
\setcounter{equation}{0}
\makeatletter
\@addtoreset{equation}{section}
\makeatother
\renewcommand{\theequation}{A.\arabic{equation}}
The five--dimensional superspace is spanned by the supercoordinates
$Z^{\underline\mu} = (x^\mu, \theta^{m i})$, where $x^\mu (\mu = 0,\ldots,4)$ are the
ordinary space--time coordinates and $\theta^{m i} (m =
1,\ldots,4)$ are symplectic--Majorana spinors carrying the $USp(2)$
doublet index $i=1,2$ which is raised and lowered with
the invariant $USp(2)$ tensor $\e_{12}= \e^{12} = 1$ as follows:
\beq
\th^{m i} = \e^{ij} \th_{m j}, \qquad \th^m_{i} = \th^{m j}
\e_{ji}.
\eeq
The flat superspace indices are $\ua=(a\,\a)$.
The vector ones are raised and lowered with the flat metric
$\eta_{ab} = \{-++++\}$.

The symplectic--Majorana condition on a generic spinor $\l_{\a i}$
reads
\beq
\ol{\l}^i \equiv \l_i^\dagger \g^0 = {}^t\l^i C,
\eeq
where $\ol{\l}$ is the usual Dirac conjugate and $C$ is the charge
conjugation matrix satisfying
$^t C = -C = C^{-1}$.

The five--dimensional ${(\g^a)_\a}^\b$ matrices satisfy the Dirac
algebra
$$
\{ \g^a, \g^b \} = 2 \, \eta^{ab}.
$$
The spinorial indices can be naturally raised and lowered through
the use of the charge conjugation matrix $C_{\a\b}$, $^t C = -C =
C^{-1}$.

To have matrices with fixed symmetry properties we define
\beq
\begin{array}{rclcrcl}
{(\G^{[n]})_{\a}}^\b  &\equiv& {(\g^{[n]})_\a}^\b, &&
{(\G^{[n]})^{\a}}_\b  &\equiv& C^{\a\rho}{(\g^{[n]})_\rho}^\s
C_{\s\b}, \\
(\G^{[n]})_{\a\b}  &\equiv& {(\g^{[n]})_\a}^\s C_{\s\b},
 &
\hbox{and } & (\G^{[n]})^{\a\b}  &\equiv& C^{\a\s}
{(\g^{[n]})_\s}^\b,
\end{array}
\eeq
where $\g^{[n]}$ means the antisymmetrized product of $n$ $\g$
matrices with weight one:
$\g^{[n]} = \g^{[a_1} \ldots \g^{a_n]}.$
Thus
\beq
^t \G^a = - \G^a, \quad ^t \G^{ab} = \G^{ab} \hbox{ and }
^t\G^{abc} = \G^{abc}.
\eeq

We define the tangent space components of a generic $p$--form
$\Phi_p$ according to \cite{BW}
$$
\Phi_p = \frac{1}{p!} e^{\ua_1} \ldots e^{\ua_p}
\Phi_{\ua_p \ldots \ua_1},
$$
where the wedge product between forms is understood.

%

\end{document}